\documentstyle[12pt]{article}
\begin{document}
\title{{\bf GRAVITATIONAL
\\ CAPTURE AND SCATTERING
\\ OF STRAIGHT TEST STRINGS
\\ WITH LARGE IMPACT PARAMETERS}
\thanks{Alberta-Thy-05-98, gr-qc/9804088}}
\author{
Don N. Page
\thanks{Internet address:
don@phys.ualberta.ca}
\\
CIAR Cosmology Program, Institute for Theoretical Physics\\
Department of Physics, University of Alberta\\
Edmonton, Alberta, Canada T6G 2J1
}
\date{(1998 April 30)}

\maketitle
\large

\begin{abstract}
\baselineskip 14.8 pt

	The capture or scattering
of an initially straight infinite test
cosmic string by a Kerr-Newman black hole,
or by any other small source
of an electrovac gravitational field,
is analyzed analytically
when the string moves with
initial velocity $v_0$ and
large impact parameter $z_0 = b \gg M$
so that the string stays very nearly straight
(except during the final capture process, if that occurs,
or except far behind the gravitating object,
if $b \not\gg M/\sqrt{1-v_0^2}\:$).
The critical impact parameter for capture
at low velocities,
$v_0 \ll 1 - Q^2/M^2$,
is shown to be
$b_{\rm{crit}} = \sqrt{(\pi/2)(M^2-Q^2)/v_0} \; + \; O(M)$.
For all $b > b_{\rm crit}$,
the displacement of the string
from the plane of the gravitating object
after the scattering approaches the final value
$z_f \approx \sqrt{b^2 - (\pi/2)(M^2-Q^2) / v_0} \; 
 - 2\pi M v_0/\sqrt{1-v_0^2}$,
for any $v_0$, so long as $b \gg M$.

\end{abstract}
\normalsize
\baselineskip 15 pt
\newpage

\section{Introduction}

	The scattering and capture of classical particles
by black holes and other gravitating objects
has been thoroughly analyzed,
but the scattering and capturing of long cosmic strings
has received much less attention \cite{Moss, JPD,JPDnew},
no doubt because the extension of the string
makes it a much more difficult problem.
In the ultrarelativistic limit of a string moving
very near the speed of light
in a direction perpendicular to the string,
the propagation of disturbances along the string
suffers a large time dilation,
so that each piece of the string is effectively decoupled
and moves very nearly along a null geodesic.
Therefore, in this limit one can easily calculate
the scattering of a string by a black hole
(at least if one assumes that loops which
may form in the string do not break off).
For example, De Villiers and Frolov \cite{JPD}
find that the critical
impact parameter for capture is the same as that
of a photon, $3\sqrt{3} \; M$ for the Schwarzschild metric.

	However, for lower velocities, disturbances
have time to propagate along the string,
and so the problem becomes much more difficult.
So far only numerical studies have been made
of the capture of these slower strings by black holes
\cite{Moss,JPD},
and, until the final part of \cite{JPDnew}, which was done
independently of this present paper
and completed just shortly after my first rough draft,
no one had found how the scattering behaves
in the limit of very low velocity.

	In this paper I shall show that the scattering
of test strings can be calculated
analytically when the strings stay sufficiently far
from the gravitating object
that they remain nearly straight.
One finds that at sufficiently low velocities
(depending upon the impact parameter),
the scattering is dominated by terms quadratic
in the mass and charge of the gravitating object.
One can easily estimate the dependence
of the critical impact parameter
on the initial string velocity
(going as its inverse square root).
By a simple calculation that was
a few orders of magnitude more difficult for me,
one can also explicitly get
the precise numerical coefficient.

\section{Power-law velocity dependence of the critical impact parameter}

	First, let us get the power-law velocity
dependence by a simple argument
that does not require writing down any equations.

	When the string is moving very slowly,
then a long portion of it that is nearest the gravitating object
will follow a sequence of nearly static configurations.
Make the assumption that the string
is everywhere far from the gravitating object compared
with the gravitating object's intrinsic size
(which for a black hole
is of the order of the hole mass $M$,
using units here and throughout this paper
with $G = c = 4\pi\epsilon_0 = 1$,
since the charge $Q$
and angular momentum $J = Ma$
merely give fractional corrections
to the size that are of order unity).
Then at the location of the string,
the Kerr-Newman (rotating, charged) black hole metric,
or the metric of any gravitational source
that is small compared with the distance to the string,
can be adequately approximated
by the static spherically symmetric Reissner-Nordstrom metric
with the same mass $M$ and charge $Q$.

	We shall consider here only a Nambu-Goto test string
(i.e., a relativistic string with tension equal to its energy
per unit length, $\mu$, in the frame of the string,
both of which are negligibly small
compared to all other relevant parameters
in order that the string be a test string
which does not affect the background spacetime geometry
 --- this requires that $\mu \ll M \gamma_0 v_0/b$ \cite{Pagenon},
where $M$ is the mass of the gravitating object,
$v_0$ is the initial velocity of the string in the frame
of the object,
$\gamma_0 \equiv 1/\sqrt{1-v_0^2}$
is the relativistic gamma factor,
and $b$ is the string impact parameter).
Each static configuration of a Nambu-Goto test string
in a static metric has a spatial configuration
that follows a geodesic of an auxiliary
spatial metric that is obtained from
the spatial part of the actual spacetime metric
by multiplying it by $-g_{00}$,
the absolute square of the magnitude
of the time translation Killing vector \cite{who1}.
In the weak-field limit of a static metric
with a nonrelativistic source
(i.e., one with pressure much less than the
energy density, including the rest mass
energy density in the latter),
the source of the linearized metric perturbation
from the Minkowski flat spacetime metric
is mainly the energy density $T_{00}$,
and this gives rise to the trace-reversed
metric perturbation $\bar{h}_{\mu\nu}$
that is dominated by the $\bar{h}_{00}$ term.
When one reverses the trace of this to get
the regular metric perturbation $h_{\mu\nu}$
and multiplies the resulting spatial part
of the metric by $-g_{00}$,
one gets an auxiliary spatial metric
that is flat to first order in the dominant metric perturbations
$h_{\mu\nu}$ that are generated by
the dominant stress-energy component $T_{00}$.
Hence a static string undergoes no bending
to first order in a static weak-field metric
of a nonrelativistic source \cite{Vil81}.

	By Birkhoff's theorem,
the spherically symmetric vacuum
(or electrovac, including the stress-energy tensor
of a spherically symmetric electromagnetic field,
when the charge $Q$ is nonzero)
solution of Einstein's equations
is the one-parameter ($M$) static vacuum
Schwarzschild metric when $Q=0$,
or the two-parameter ($M$ and $Q$) static
electrovac Reissner-Nordstrom metric \cite{MTW}.
Hence in the general spherically symmetric case
in the weak-field limit
(to which the general stationary Kerr-Newman
metric is an adequate approximation
at large distances where the field is weak),
one gets the same result as for
a weak-field static metric
with a nonrelativistic source:
static strings do not bend to first order in $M$,
even for a black hole.

	(Static neutral strings also do not bend to first order in $Q$,
since the electromagnetic stress-energy tensor
and Reissner-Nordstrom metric depend quadratically on $Q$.
Like $M$, $Q$ has the dimension of length when
$G = c = 4\pi\epsilon_0 = 1$
and has a magnitude no larger than $M$ for a black hole.)

	One can also see this result directly for
the Reissner-Nordstrom metric
when one multiplies its spatial part by $-g_{00}$
and replaces the usual radial variable $r$
with a new radial variable that is $r-M$,
since the resulting auxiliary spatial metric,
where a static string follows a geodesic,
is flat except for a correction term proportional
to $M^2-Q^2$.

	Therefore, the quasistatic sequence of string
configurations that pass by at great distance
and low velocity from a black hole
or other small gravitational source
are not bent to first order in $M$,
but they are to second order
(actually by an amount proportional to
the correction term in the auxiliary spatial metric
that is proportional to $M^2-Q^2$).
If the string is not captured,
its nearest distance from the gravitating object
will be of the order of its impact parameter
$b=z_0$ (the displacement of the plane
of the string's initial motion from the gravitating object).
When the nearest distance of the string from
the gravitating object is of order $b$,
by dimensional analysis it is obvious that
the bending angle is proportional to
$M^2/b^2$ (more precisely, to $(M^2-Q^2)/b^2$,
but for simplicity I shall assume that
$M^2-Q^2$ is of the same order as $M^2$,
which it will be for a black hole unless it
is very near an extremal Reissner-Nordstrom
black hole that has $Q^2 = M^2$,
and which it will be in for a general gravitating object
if the charge $Q$ of the object is not very large
in comparison with its total mass-energy $M$).

	The actual displacement of the string
configurations from the straight ones
they would follow in Minkowski spacetime
in the absence of the gravitational field
depends on how great a length of it
is bent by an angle of order $M^2/b^2$.
This will be the length along the string
for which there has been time for causal influences
to propagate (i.e., for the string to bend)
during the time that the string passes
at a distance of order $b$ from the gravitating object.
Since the string is traveling at a speed
of the order of its initial speed $v_0$,
it takes a time of order $b/v_0$ to pass
the region where it is of order its nearest
distance from the gravitating object.
With the bending perturbation
traveling at nearly unit speed ($c=1$)
along the slowly moving string,
a length of string of order $b/v_0$
is bent by an angle of order $M^2/b^2$,
giving a displacement of the bent
part of the string of order $M^2/(b v_0)$.

	Assuming that the displacement
is toward the gravitating object
(which it is, when $M^2-Q^2>0$, as I shall generally assume),
the string will not be captured if its impact parameter
$b$ is greater than this bending displacement,
which will be the case if $b^2$ is greater than
some quantity of the order of $M^2/v_0$.
However, the string will generally be captured
(certainly in the black hole case)
if the string bends down to reach the gravitating object,
which we would expect to be the case
when $b^2$ is smaller.
Thus we can estimate the critical impact parameter
for slow string capture to be of the order of $M/\sqrt{v_0}$
(or, more precisely, proportional to $\sqrt{(M^2-Q^2)/v_0}$
in the limit that the initial string speed $v_0$ is sufficiently small).	
	This part of the argument can easily
be done in one's head in a few moments of quiet reflection
(or in semi-quiet reflection, as in washing dishes).
Now comes the dirty work of finding the precise
numerical coefficient, which took me about a month of algebra
(between various other duties,
including the previously-mentioned moments
of semi-quiet reflection),
if one counts my various false starts.

\section{Outline of the method and approximations}

	The method for calculating the capture and scattering
of infinitely long, nearly straight, test cosmic strings
with large impact parameter $b$
will be to write down approximate
expressions for the string motion in two overlapping
domains around the gravitating object.
In the inner domain,
the string configuration at each moment of time
is assumed to be very nearly straight
and thus determined by the position of its
nearest point to the gravitating object.
In this domain one calculates the small bending
of the string to first order in deviations
of the metric and connection from flatness,
evaluated at the zeroth order position
of the nearly straight string.
In the outer domain,
at distances much larger than
the impact parameter $b$ from the gravitating object,
the metric is sufficiently
flat that its effect on the string motion
is negligibly different from that of Minkowski spacetime,
and one simply has outgoing perturbations
of the initially straight string configuration.
The small bending of the string by the nonflat metric
in the inner domain is taken to be the source
of these outgoing perturbations in the
approximately flat outer domain.

	This approximation method only works
when the string stays nearly straight
and is not bent at large angles
by the metric of the gravitating object,
a condition which is satisfied when the string
is always far from the object compared with the
object size, which is of order $M$ for a black hole.
(More strictly correct, for a string
with an arbitrary initial relativistic gamma-factor
$\gamma_0 = 1/\sqrt{1-v_0^2}$,
when the impact parameter $b = z_0$
is large compared with the energy
$M\gamma_0$ of the gravitating object in the initial string frame,
since for $\gamma_0 \geq b/(2M) \gg 1$
one gets cusps and loops in the string \cite{Page_fast}.
Nevertheless, the cusps and loops develop
only far behind the gravitating object,
where it focuses the different pieces
of the string to cross,
and so even for large $M\gamma_0/b$
the string stays nearly straight
while it is having a significant
gravitational interaction with the object.)

	However, the method does not require that
the configuration of the string stay spatially near
what its unperturbed motion would be
if the gravitating object were absent:
the point on the string that at each moment
of time is nearest to the gravitating object
can follow a trajectory that has large
deviations from the straight line
it would follow at constant velocity
if the string were unperturbed.
The only assumption is that at each moment
the string configuration passing through this point
(where the string is nearest the gravitating object)
is nearly straight,
with small bending perturbations
(small in angle, not necessarily small
in the total linear deviation from the unperturbed
straight string configurations)
propagating outward along the string
at the speed of light in the nearly flat
outer domain.
In this way the range of applicability
of the calculations presented here
extends to lower initial velocities,
$v_0 \not\gg (M^2-Q^2)/b^2$
(in particular, to such low velocities
that the string is captured),
than the results obtained by
a different method of calculation
by De Villiers and Frolov \cite{JPDnew}.

	Our results are in perfect agreement
in the overlap region
$(M^2-Q^2)/b^2 \ll \gamma_0 v_0 \ll b/M$,
where $\gamma_0 v_0 \equiv v_0/\sqrt{1-v_0^2}$,
the magnitude of the spatial component
of the four-velocity,
is what I propose be named the celerity
(the magnitude of the spatial momentum per rest mass,
i.e., the hyperbolic sine of the rapidity,
as the ordinary three-velocity magnitude or speed
is the hyperbolic tangent of the rapidity,
and the energy per rest mass, $\gamma_0$ itself,
is the hyperbolic cosine of the rapidity).
In this overlap region where both methods work,
the string is only slightly scattered
and is nowhere near to being captured
(which it would do at lower celerities,
where the left hand inequality is not satisfied)
and is also nowhere near to forming loops
behind the string
(which it would do at higher celerities \cite{Page_fast},
where the right hand inequality is not satisfied).

	In both spatial domains in my method of calculation
(the domain closer to the gravitating object,
where the string is at each moment of time
nearly where it would be if it were straight,
and the domain farther away,
where one can take spacetime to be flat),
I am assuming that the string
stays sufficiently far from the gravitating object
(except for its final capture, which will not be analyzed here,
only the conditions that lead to its being captured)
that it is only slightly bent, and that it is adequate
to consider the gravitating object metric only to quadratic
order in its mass $M$ and charge $Q$,
i.e., only to order $M^2/r^2$ and $Q^2/r^2$
for these dimensionless quantities.
I shall only consider these quadratic terms when
they persist at $v_0 = 0$, since otherwise
they are dominated by terms that are first order in $M/r$
(which all are multiplied by factors that go to zero
linearly in $v_0$).
In other words, I shall consider only terms that
are of the lowest nontrivial order
in $M/r$ (or in $Q/r$, which is counted as being
of the same order; the lowest nontrivial order
of both of these is quadratic when $v_0$ is very small)
and in $Mv_0/r$.

	One might also ask about the effect of the angular
momentum $J=Ma$ of the gravitating object.
It has the dimension of squared length
and is no greater than $M^2$
for a Kerr-Newman gravitating object, so I shall count it
as potentially of order $M^2$
(like $Q^2$,
since neither can be larger than $M^2$,
though of course either or both could be much smaller,
but I need not assume that either or both are much smaller).
The angular momentum couples linearly
to the velocity of the string at low velocities
to give dimensionless terms
of order $J v_0/r^2$ or $M^2 v_0/r^2$,
but since these terms are taken to be
negligible compared with the $M v_0/r$
terms I am retaining,
this coupling can be neglected
in the approximations I am making.

	There is also a velocity-independent effect
(more strictly, an effect that persists
when the string velocity is taken to zero)
of the angular momentum upon the string motion
(upon the quasistatic configurations of the strings \cite{who1}),
but this starts quadratically in $J$ and hence gives
dimensionless effects proportional to $J^2/r^4$,
which is of order $M^4/r^4$ and hence will be neglected.
Using the upper bound on the angular momentum
of a Kerr-Newman black hole, one can see that $J^2/r^4$
is less than or equal to $M^2(M^2-Q^2)/r^4$
and hence is negligible compared with
the $(M^2-Q^2)/r^2$ term I am keeping,
when the string is at $r\gg M$ as I am assuming
in order that the string bending be small.
For other gravitating objects in which $J^2/r^4$
is not bounded thus, I am assuming
that it is still negligible compared with $(M^2-Q^2)/r^2$.

	The net result of this consideration
of the angular momentum of the gravitating object is that
I can neglect it for my considerations
of strings propagating far from the gravitating object.
(If the angular momentum is of order $M^2$,
its effects would not be negligible
on strings propagating near the gravitational radius
of the gravitating object,
but I am not considering such strings in this paper.)

	If the gravitating object is not
a stationary electrovac black hole
and so can have multipole moments that are not
fixed by its mass, charge, and angular momentum,
I shall assume that the string stays far enough
away from the object, in comparison with the object size,
that one can neglect the effect of all of these
multipole moments.
For a cosmic string with $\mu \sim 10^{-6}$,
which is not actually infinitesimal,
it would be generally be problematic to 
stay sufficiently far from a nonrelativistic
gravitating object
and remain in the test string approximation,
$\mu \ll M \gamma_0 v_0/b$ \cite{Pagenon},
so in practice for such cosmic strings the test string
approximation adopted in this paper
would be adequate for capture or scattering only by
black holes or other very compact relativistic objects,
such as neutron stars.
However, in this paper I shall continue to assume
that the test string approximation is good,
and that the string stays far enough from
the gravitating object that its angular momentum
and multipole moments have a negligible effect
upon the motion of the string.

\section{Derivation of the equations of motion
of the string location}

	As a result of the assumptions
and approximations above,
it is an adequate approximation
for my present purposes to use the
unique spherically symmetric electrovac
metric with the same mass $M$ and charge $Q$
as that of the actual gravitating object,
namely, the Reissner-Nordstrom metric
 \begin{equation}
 ds^2 = - \left( 1-{2M\over r}-{Q^2\over r^2} \right) dt^2
		+ \left( 1-{2M\over r}-{Q^2\over r^2} \right)^{-1} dr^2
		+ r^2 d\theta^2
		+ r^2 \sin^2{\theta}d\varphi^2.
 \label{eq:1}
 \end{equation}

	Although this is a slight aside from the approach
I shall actually use, it is instructive to consider
the precise bending angle for all static string configurations
in this metric.  As mentioned above,
these follow geodesics of the auxiliary spatial metric
obtained by multiplying the spatial part of the static metric
by $-g_{00}$.  Or, for a stationary metric
 \begin{equation}
 ds^2 = g_{00}(x^k) dt^2 + 2 g_{0i}(x^k) dt dx^i +g_{ij}(x^k) dx^i dx^j,
 \label{eq:2}
 \end{equation}
such as the Kerr-Newman metric,
the auxiliary spatial metric,
wherein static strings
(strings which have their spatial coordinates $x^k$
stay fixed as the time coordinate $t=x^0$ evolves)
follow geodesics, is \cite{who1}
 \begin{equation}
 ds_{\rm aux}^2 = (g_{0i} g_{0j} - g_{00} g_{ij}) dx^i dx^j.
 \label{eq:3}
 \end{equation}
However, the auxiliary spatial metric
for the stationary Kerr-Newman metric differs
(in the appropriate coordinate system)
from the auxiliary spatial metric
for the static Reissner-Nordstrom metric
only by terms that are quadratic
in the angular momentum $J$
and hence which are negligible at the
distant locations at which I am generally
considering the strings to move.

	For the Reissner-Nordstrom metric (\ref{eq:1})
above, the auxiliary spatial metric  (\ref{eq:3}) takes the form
 \begin{equation}
 ds_{\rm aux}^2 = dR^2
	 + (R^2 - M^2 +Q^2)(d\theta^2 + \sin^2{\theta}d\varphi^2),
 \label{eq:4}
 \end{equation}
where, to show more explicitly that this is
flat up to the correction terms proportional to $M^2-Q^2$,
I have shifted the radial coordinate to $R = r-M$.
If a static string, which follows a geodesic
of this auxiliary spatial metric (\ref{eq:4}),
has its minimum value of $R$ as $R_0$,
then the bending angle is \cite{who1}
 \begin{equation}
 2\delta = \pi - 2k'K(k)
 = \pi\left( {1\over 4}k^2 + {7\over 64}k^4 + {17\over 256}k^6 + O(k^8) \right)
 \approx {\pi (M^2-Q^2) \over 4 R_0^2},
 \label{eq:5}
 \end{equation}
where $K(k)$ is the complete elliptic integral
of the first kind with modulus
 \begin{equation}
 k = {\sqrt{M^2-Q^2}\over R_0}
 \label{eq:6}
 \end{equation}
and complementary modulus
 \begin{equation}
 k' = \sqrt{1-k^2} = {\sqrt{R_0^2 - M^2 + Q^2}\over R_0}.
 \label{eq:7}
 \end{equation}

	Now let us return to the question of what the motion
of a nearly straight string is far from a gravitating object,
in a metric that at that great distance
is adequately approximated
by the Reissner-Nordstrom metric.
In a general curved spacetime,
the motion of a free infinitesimally thin
relativistic cosmic test string
with the Nambu-Goto action \cite{Nambu,Goto}
(proportional to the area of the worldsheet
swept out by the string)
obeys the equation of motion \cite{Carter}
 \begin{equation}
 P^{\alpha}_{\;\;\;\beta}P^{\beta}_{\;\;\;\gamma;\alpha} = 0,
 \label{eq:8}
 \end{equation}
where $P^{\alpha}_{\;\;\;\beta}$
is the rank-two projection tensor
into the tangent space of the string worldsheet.
This is actually the form of the equation
of motion of any free $n$-brane in curved spacetime
\cite{Carter},
such as the 0-brane or free particle,
for which Eq. (\ref{eq:8})
becomes the geodesic equation with then
$P^{\alpha}_{\;\;\;\beta} = - u^{\alpha} u_{\beta}$
being the rank-one projection tensor
to the tangent space of the particle worldline
(the space of vectors parallel to the
particle four-velocity $u^{\alpha}$).

	Although Eq. (\ref{eq:8}) is written
purely in terms of structures in the spacetime
and does not require any parametrization
of the string worldsheet,
for my analysis below it is convenient
to introduce null coordinates $(u,v)$
on the worldsheet itself,
so that with coordinates $x^{\alpha}$
for the spacetime, the string worldsheet
takes the form $x^{\alpha} = x^{\alpha}(u,v)$.
In order that the coordinates $(u,v)$
be indeed null in the induced geometry
on the string worldsheet,
one must have the constraint equations
 \begin{equation}
 g_{\alpha\beta} x^{\alpha}_{,u} x^{\beta}_{,u}
 = g_{\alpha\beta} x^{\alpha}_{,v} x^{\beta}_{,v} = 0,
 \label{eq:9}
 \end{equation}
where a comma denotes a partial derivative.
Then Eq. (\ref{eq:8}) becomes
 \begin{equation}
 x^{\alpha}_{,uv}
 + \Gamma^{\alpha}_{\;\beta\gamma}
	x^{\beta}_{,u}x^{\gamma}_{,v} = 0,
 \label{eq:9b}
 \end{equation}
where $\Gamma^{\alpha}_{\;\beta\gamma}$
are the Christoffel symbols of the spacetime
metric in the basis of spacetime coordinates $x^{\alpha}$.

	It will be convenient to use spatial coordinates
$x \equiv x^1$, $y\equiv x^2$, and $z \equiv x^3$,
with $\rho \equiv (x^2 + y^2 + z^2)^{1/2}$
being a new radial variable
(and with the superscripts in this last equation
being exponents, unlike in the previous three
equations in this sentence, where they
are particular choices of the indices $i$
used to denote spatial indices $x^i$),
and to choose these coordinates
so that the spatial part of the
Reissner-Nordstrom metric (\ref{eq:1})
is manifestly conformally flat.
This requires that the previously used
radial coordinates $r$ and $R$ be given by
 \begin{equation}
 R = r-M = \rho + {M^2-Q^2\over 4\rho}.
 \label{eq:9c}
 \end{equation}

	In these new coordinates,
the Reissner-Nordstrom metric takes the form
 \begin{eqnarray}
 ds^2 &=& - e^{2\phi + 2\psi} dt^2
		+ e^{-2\phi} (dx^2 + dy^2 + dz^2)
	\nonumber \\
	&=& \left( 1 + {M\over\rho} + {M^2-Q^2\over 4\rho^2} \right)^{-2}
		\left( 1 - {M^2-Q^2\over 4\rho^2} \right)^2 dt^2
	\nonumber \\
	&+& \left( 1 + {M\over\rho} + {M^2-Q^2\over 4\rho^2} \right)^2
		(dx^2 + dy^2 + dz^2)
 \label{eq:10}
 \end{eqnarray}
with
 \begin{equation}
 \phi = - \ln{\left( 1 + {M\over\rho} + {M^2-Q^2\over 4\rho^2} \right)}
	\approx - {M \over \rho},
 \label{eq:11}
 \end{equation}
 \begin{equation}
 \psi = \ln{\left( 1 - {M^2-Q^2\over 4\rho^2} \right)}
	\approx - {M^2-Q^2\over 4\rho^2}.
 \label{eq:12}
 \end{equation}

	When $M = Q = 0$, this metric is obviously
the flat Minkowski metric,
and in it I shall take the unperturbed solution
to be a straight string extended in the $x$-direction,
moving with speed $v_0$ in the $y$-direction,
and at impact parameter or height $z=b$
above the $x-y$ plane.
With a suitable choice of null coordinates
$(u,v)$ on the string worldsheet
(each of which can be replaced
by an arbitrary smooth monotonic function
of itself to give a new null coordinate),
the location of this unperturbed string worldsheet
in the flat Minkowski metric can be given by
 \begin{eqnarray}
 t &=& t_0 \ \equiv \ {1\over 2} (u+v), \label{eq:13} \\
 x &=& x_0 \ \equiv \ {1\over 2} \gamma_0^{-1} (v-u),
	\label{eq:14} \\
 y &=& y_0 \ \equiv \ {1\over 2} v_0 (u+v),  \label{eq:16} \\
 z &=& z_0 \ \equiv \ b.  \label{eq:17}
 \end{eqnarray}
Here $\gamma_0 = 1/\sqrt{1-v_0^2}$
is the usual relativistic gamma-factor.
Note that the (constant) speed $v_0$
is not to be confused with the null coordinate $v$.

	Now when $M$ and possibly also $Q$ are nonzero,
I shall take the string worldsheet to start off with the form
of Eqs. (\ref{eq:13}) - (\ref{eq:17}) at the infinite past,
at $t_0 = - \infty$ for all $x_0$.
Then in the outer domain, $\rho \gg b$,
where the spacetime curvature may be ignored in comparison
with the curvature in the region $\rho \sim b$
that the part of the string with
$|x_0| {\ \lower-1.2pt\vbox{\hbox{\rlap{$<$}
\lower5pt\vbox{\hbox{$\!\!\sim$}}}}\ } b$
reaches at
$|t_0| {\ \lower-1.2pt\vbox{\hbox{\rlap{$<$}
\lower5pt\vbox{\hbox{$\!\!\sim$}}}}\ } b/v_0$,
there will essentially just be outgoing
flat-spacetime perturbations of the string for
$u {\ \lower-1.2pt\vbox{\hbox{\rlap{$>$}
\lower5pt\vbox{\hbox{$\!\!\sim$}}}}\ } - b/v_0$,
$v {\ \lower-1.2pt\vbox{\hbox{\rlap{$>$}
\lower5pt\vbox{\hbox{$\!\!\sim$}}}}\ } - b/v_0$.

	In particular,
the string worldsheet will have in the outer domain
the approximate form
 \begin{eqnarray}
 t &\approx& t_0 + \theta(v-u) T(u) + \theta(u-v) T(v), \label{eq:18} \\
 x &\approx& x_0 + \theta(v-u) X(u) - \theta(u-v) X(v), \label{eq:19} \\
 y &\approx& y_0 + \theta(v-u) Y(u) + \theta(u-v) Y(v),  \label{eq:20} \\
 z &\approx& z_0 + \theta(v-u) Z(u) + \theta(u-v) Z(v),  \label{eq:21}
 \end{eqnarray}
where I have used the fact that the perturbations from
the metric in the region $\rho \sim b$, and the string worldsheet,
are symmetric under $x\rightarrow -x$,
or under $u\leftrightarrow v$ on the string worldsheet.
Here $\theta$ is the standard Heaviside step function,
0 for negative argument, 1/2 for argument zero,
and 1 for positive argument.
Therefore, the right half of the string with $x > 0$ ($v > u$)
will have right-moving perturbations that are functions
of $u$ only and which move at the speed of light rightward
away from the region $\rho \sim b$ where they are generated,
and the left half of the string with $x < 0$ ($u > v$)
will have left-moving perturbations that are functions
of $v$ only and which move at the speed of light leftward
away from the region $\rho \sim b$ where they are generated.

	In the inner domain $\rho \sim b$ where the metric
deviations from flat spacetime generate the perturbations
(bending) of the string, Eqs. (\ref{eq:18}) - (\ref{eq:21})
are not valid in its separation of the perturbations
into left- and right-moving modes only.
However, so long as $\rho \gg M$,
as we have been assuming,
the string will be sufficiently straight in this inner domain
that to an excellent approximation
we may calculate the metric deviations that generate
the string perturbations
as if they acted at the location of a straight string
passing though the trajectory that follows the $x=0$
midpoint of the string, where $u=v=t_0$.
Matching this straight string
in the inner domain with Eqs. (\ref{eq:18}) - (\ref{eq:21})
in the outer domain gives in the inner domain
 \begin{eqnarray}
 t &\approx& t(t_0) \ = \ t_0 + T(t_0), \label{eq:22} \\
 x &\approx& (1 - \gamma_0 \dot{X}(t_0)) x_0
	 \ \equiv \ {1\over 2}\gamma^{-1}(v-u),
	 \label{eq:23} \\
 y &\approx& y(t_0) \ = \ y_0 +Y(t_0) \ = \ v_0 t_0 + Y(t_0),  \label{eq:24} \\
 z &\approx& z(t_0) \ = \ z_0 + Z(t_0) \ = \ b + Z(t_0)  \label{eq:25}
 \end{eqnarray}
Here and henceforth, an overdot will denote a derivative
with respect to $t_0 \equiv (u+v)/2$ at $x=0$,
where $t_0 = u = v$.
In particular, $\dot{t} = 1 + \dot{T}$, $\dot{y} = v_0 + \dot{Y}$,
and $\dot{z} = \dot{Z}$.

	From the constraint equations (\ref{eq:9})
that $u$ and $v$ be null coordinates on the string worldsheet,
one can see that, taking the metric to be approximately Minkowskian,
 \begin{equation}
 \gamma \ = \ \gamma(t_0) 
	\ \approx \ (\dot{t}^2 - \dot{y}^2 - \dot{z}^2)^{-1/2}.
 \label{eq:26}
 \end{equation}

	As a results of these approximations that the string
is nearly straight in the inner domain
and takes the flat spacetime form with outgoing
perturbations in the outer domain, determining
the approximate configuration of the string worldline
reduces to finding the three functions of one variable
$T$, $Y$, and $Z$, or, equivalently,
$t(t_0)$, $y(t_0)$, and $z(t_0)$.
In order to get ordinary differential equations for them,
use the string equations of motion in the coordinate
form of Eq. (\ref{eq:9b}), integrated over $x$
at fixed $t$.

	Now using the approximate form of the string worldsheet
as seen in the large-scale view from the outer domain,
given by Eqs. (\ref{eq:18}) - (\ref{eq:21}),
one can integrate the various $x^{\alpha}_{,uv}$
in Eq. (\ref{eq:9b})
over $x$ at fixed $t$ to get
 \begin{eqnarray}
 \int{dx \: t_{,uv}} &\approx&{1\over 2\gamma} \: \dot{T}
	\ = \ {1\over 2\gamma} \: (\dot{t} - 1), \label{eq:27} \\
 \int{dx \: x_{,uv}} &\approx& 0, \label{eq:28} \\
 \int{dx \: y_{,uv}} &\approx&{1\over 2\gamma} \: \dot{Y}
	\ = \ {1\over 2\gamma} \: (\dot{y} - v_0),  \label{eq:29} \\
 \int{dx \: z_{,uv}} &\approx&{1\over 2\gamma} \: \dot{Z}
	\ = \ {1\over 2\gamma} \: \dot{z}.  \label{eq:30}
 \end{eqnarray}

	To integrate the various
$\Gamma^{\alpha}_{\;\beta\gamma} x^{\beta}_{,u}x^{\gamma}_{,v}$
in Eq. (\ref{eq:9b}) over $x$ at fixed $t$,
use the approximately straight form of the string
in the inner region for evaluating the $u$- and $v$-derivatives
of $t(u,v)$, $y(u,v)$, and $z(u,v)$
as functions purely of $t_0 = (u+v)/2$
in terms of how $t$, $y$, and $z$
behave at $x=0$ as functions of $t_0$.
However, I shall use the constraint equations (\ref{eq:9})
to determine the  $u$- and $v$-derivatives of $x(u,v)$.
In particular, I take
 \begin{eqnarray}
 t_{,u} &=& t_{,v} \ \approx \ {1\over 2} \: \dot{t}, \label{eq:31} \\
 x_{,u} &=& - x_{,v} \ \approx \ 
 - \, {1\over 2} \: \sqrt{e^{4\phi+2\psi} \: \dot{t}^2 - \dot{y}^2 - \dot{z}^2},
	 \label{eq:32} \\
 y_{,u} &=& y_{,v} \ \approx \ {1\over 2} \: \dot{y},  \label{eq:33} \\
 z_{,u} &=& z_{,v} \ \approx \ {1\over 2} \: \dot{z}.  \label{eq:34}
 \end{eqnarray}

	Then one gets, using a prime on the exponents
$\phi$ and $\psi$ in the first form of the metric (\ref{eq:10})
to denote a derivative with respect to the radial
coordinate $\rho$,
 \begin{eqnarray}
 \Gamma^0_{\;\beta\gamma} x^{\beta}_{,u} x^{\gamma}_{,v}
 &\approx& {\phi' + \psi' \over 2\rho} \: \dot{t} \: (y\dot{y} + z\dot{z}),
	 \label{eq:35} \\
 \Gamma^1_{\;\beta\gamma} x^{\beta}_{,u}x^{\gamma}_{,v}
 &\approx& e^{4\phi+2\psi}\: {2\phi' + \psi' \over 4\rho} \: \dot{t}^2 \: x,
	 \label{eq:36} \\
 \Gamma^2_{\;\beta\gamma} x^{\beta}_{,u}x^{\gamma}_{,v}
 &\approx& e^{4\phi+2\psi}\: {\psi' \over 4\rho} \: \dot{t}^2 \: y
	- {\phi' \over 2\rho} \: \dot{z} \: (z\dot{y} - y\dot{z}),
	 \label{eq:37} \\
 \Gamma^3_{\;\beta\gamma} x^{\beta}_{,u}x^{\gamma}_{,v}
 &\approx& e^{4\phi+2\psi}\: {\psi' \over 4\rho} \: \dot{t}^2 \: z
	+ {\phi' \over 2\rho} \: \dot{y} \: (z\dot{y} - y\dot{z}).
	 \label{eq:38}
 \end{eqnarray}
These expressions apply at any radius
for any static spherically symmetric metric,
which can be put into the form
of the first form of the metric (\ref{eq:10}).
The only approximation is the approximation
of Eqs. (\ref{eq:31}) - (\ref{eq:34})
for the $u$- and $v$-derivatives of the
spacetime coordinates ($t$, $x$, $y$, $z$)
on the string worldsheet as functions
of the null coordinates $(u,v)$
on the string worldsheet.

	Now we shall explicitly go to the Reissner-Nordstrom
form of the metric (\ref{eq:10})
and assume that we are at $\rho \gg M > |Q|$.
Then using the approximate expressions in
Eqs. (\ref{eq:11}) and (\ref{eq:12})
for $\phi$ and $\psi$,
we can set $e^{4\phi+2\psi} \approx 1$
and
 \begin{eqnarray}
 {\phi' \over \rho} &\approx& {M \over \rho^3}, \label{eq:39} \\
 {\psi' \over \rho} &\approx& {M^2 - Q^2 \over 2\rho^4}.
	 \label{eq:40}
 \end{eqnarray}
Since $\rho \gg M$, we have $\phi' \gg \psi'$,
so we shall ignore $\psi'$ in any term where it is added
to a nonzero multiple of $\phi'$ with the same
factors of $\dot{t}$, $\dot{y}$, and $\dot{z}$. 

	Just as we integrated the various $x^{\alpha}_{,uv}$
in Eq. (\ref{eq:9b}) over $x$ at fixed $t$ to get
Eqs. (\ref{eq:27}) - (\ref{eq:30}),
now we can use Eqs. (\ref{eq:39}) and (\ref{eq:41})
to integrate the various
$\Gamma^{\alpha}_{\;\beta\gamma} x^{\beta}_{,u}x^{\gamma}_{,v}$
in Eq. (\ref{eq:9b}) over $x$ at fixed $t$.
If we now define
 \begin{equation}
 r \equiv \sqrt{y^2 + z^2} = \sqrt{\rho^2 - x^2},
 \label{eq:41}
 \end{equation}
the approximate distance from the gravitating object
to the nearest point (at $x=0$) on the string
(not quite the same as the previous
use of $r$ to denote a Schwarzschildean
or circumferential radial coordinate,
though to the approximations being
made, the two definitions are virtually indistinguishable
at the point on the string nearest the gravitating object), 
we obtain
 \begin{eqnarray}
 \int{dx \: \Gamma^0_{\;\beta\gamma} x^{\beta}_{,u}x^{\gamma}_{,v}}
 &\approx& {M \over r} \: \dot{t} \: \dot{r},
 \label{eq:42} \\
 \int{dx \: \Gamma^1_{\;\beta\gamma} x^{\beta}_{,u}x^{\gamma}_{,v}}
 &\approx& 0,
 \label{eq:43} \\
 \int{dx \: \Gamma^2_{\;\beta\gamma} x^{\beta}_{,u}x^{\gamma}_{,v}}
 &\approx& {\pi(M^2-Q^2)\over 16r^3} \: \dot{t}^2 \: y
 - {M \over r^2} \: \dot{z} \: (z\dot{y} - y\dot{z}),
 \label{eq:44} \\
 \int{dx \: \Gamma^3_{\;\beta\gamma} x^{\beta}_{,u}x^{\gamma}_{,v}}
 &\approx& {\pi(M^2-Q^2)\over 16r^3} \: \dot{t}^2 \: z
 + {M \over r^2} \: \dot{y} \: (z\dot{y} - y\dot{z}).
 \label{eq:45}
 \end{eqnarray}

	Now if we combine Eqs. (\ref{eq:42}) - (\ref{eq:45})
with Eqs. (\ref{eq:27}) - (\ref{eq:30})
to get the $x$-integrated form of
the string equations of motion (\ref{eq:9b}),
we get the following ordinary differential
equations for the coordinates
$t(t_0)$, $y(t_0)$, and $z(t_0)$
of the point on the string (at $x=0$) nearest the gravitating object,
with overdots denoting ordinary derivatives
with respect to $t_0$,
with $r \equiv \sqrt{y^2 + z^2}$
being the approximate distance from
the nearest point on the string to the gravitating object,
and with $\gamma = \gamma(t_0)$ being
a relativistic gamma-factor given by Eq. (\ref{eq:26}):
 \begin{eqnarray}
 \dot{t} &\approx& 1 - {2\gamma M \over r} \: \dot{r} \: \dot{t},
 \label{eq:46} \\
 \dot{y} &\approx& v_0 - {\pi\gamma(M^2-Q^2)\over 8 r^3} \: \dot{t}^2 \: y
 + {2\gamma M \over r} \: \dot{z} \: (z\dot{y} - y\dot{z}),
 \label{eq:47} \\
 \dot{z} &\approx& - {\pi\gamma(M^2-Q^2)\over 8 r^3} \: \dot{t}^2 \: z
 - {2\gamma M \over r} \: \dot{z} \: (z\dot{y} - y\dot{z}).
 \label{eq:48}
 \end{eqnarray}

	In order for the string to remain nearly straight
for all time,
it is necessary that $r \gg \gamma M$,
so Eq. (\ref{eq:46}) implies that under this condition,
$\dot{t} \approx 1$, which can be inserted into
Eqs. (\ref{eq:47}) and (\ref{eq:48}).
Under this condition, we can also,
to an excellent approximation, replace
$\gamma(t_0)$ by the constant relativistic gamma-factor
$\gamma_0 = 1/\sqrt{1-v_0^2}$.

	The appearance of Eqs. (\ref{eq:47})
and (\ref{eq:48}) can be
further simplified by defining the length scale
 \begin{equation}
 L \equiv \sqrt{\pi (M^2-Q^2) \gamma_0 \over 8 v_0}
 \label{eq:49}
 \end{equation} 
and the dimensionless parameter
 \begin{equation}
 \alpha \equiv {2 \gamma_0 M \over L}
 \equiv \sqrt{32 M^2 \gamma_0 v_0^3 \over \pi (M^2 - Q^2)} 
 \label{eq:50}
 \end{equation} 
and by replacing the independent variable
$t_0$ with $y_0 = v_0 t_0$.
Having dispensed with the radial derivatives
$\phi'$ and $\psi'$,
we can now redefine the prime
to mean differentiation with respect to $y_0$.
Then Eqs. (\ref{eq:47}) and (\ref{eq:48})
take on the simplified forms
 \begin{eqnarray}
 y' &\approx& 1 - {L^2 y \over r^3}
 + {\alpha L \over r^2} \: z' \: (zy' - yz'),
 \label{eq:51} \\
 z' &\approx& - {L^2 z \over r^3}
 + {\alpha L \over r^2} \: y' \: (zy' - yz'),
 \label{eq:52}
 \end{eqnarray}
with the initial conditions being
that at $y_0 = - \infty$, we have
$y = - \infty$ and $z = b$.

	Alternatively, one can convert these
equations into polar coordinates
such that $y = r \cos{\theta}$ and $z = r \sin{\theta}$:
 \begin{eqnarray}
 r' &\approx& \cos{\theta} - {L^2 \over r^2}
 - \alpha L r \theta'^2,
 \label{eq:53} \\
 \theta' &\approx& - {\sin{\theta} \over r}
 + {\alpha L \over r} r' \theta'.
 \label{eq:54}
 \end{eqnarray}
Another convenient pair of variables is $r$ and
$c \equiv \cos{\theta}$,
in which the equations take the simple form
 \begin{eqnarray}
 r' &\approx& c - {L^2 \over r^2}
 + {\alpha L r c'^2 \over 1-c^2},
 \label{eq:55} \\
 c' &\approx& - {1-c^2 \over r}
 + {\alpha L \over r} r' c'.
 \label{eq:56}
 \end{eqnarray}
Yet another pair, which is particularly
useful for finding the solution when
$\alpha \ll 1$, is $z$ and $c$.
For these the equations of evolution are,
using $s \equiv \sin{\theta} = \sqrt{1-c^2}$,
 \begin{eqnarray}
 z' &\approx& - {L^2 s^3 \over z^2}
 \left[ 1 + {\alpha z^2 \over L s^7} \: c' \: (s^2 c z' + z c') \right],
 \label{eq:57} \\
 c' &\approx& - {s^3 \over z}
 \left[ 1 + {\alpha L \over s^6} \: c' \: (s^2 z' + z c c') \right].
 \label{eq:58}
 \end{eqnarray}

	One can see from these equations
that the only fixed point
is at $(y,z) = (L,0)$ or $(r,\theta) = (L,0)$
or $(r,c) = (L,1)$ or $(z,c) = (0,1)$.
This represents a string
that is marginally captured by the gravitating object.
Its nearest point to the gravitating object, at $(x,y,z) = (0,L,0)$,
stays fixed in an unstable balance between
the attractive force of the gravitating object in the negative $y$-direction
toward the gravitating object location at the origin $(0,0,0)$
and the pull of the distant part of the string
(at large $|x|$)
toward the positive $y$-direction in which
the distant part of the string continues moving
at speed $v_0$.

	We can provide a partial check
on the approximations used so far
by using the precise bending angle (\ref{eq:5})
for a static string in the Reissner-Nordstrom
metric to calculate the relation between
the point of nearest approach of a marginally
captured string and the asymptotic velocity $v_0$.
Looking at the string
from a scale very large compared with $L$
(where the metric is essentially flat),
the string will have a bend of $2\delta$
near the gravitating object,
given by Eq. (\ref{eq:5}) as a function of
the coordinate $R_0$ of closest approach,
and two kinks of $\delta$ each
moving off at the speed of light
into the part of the string that is
parallel to the $x$-axis and moving
in the $y$-direction with speed $v_0$.

	By simple geometry, one can see that
if the kinks are moving away from the gravitating object
at the speed of light (one in our units)
at an angle $\delta$ from the $x$-axis,
then the component of their velocity
in the $y$-direction is $\sin{\delta}$,
which must be the same as $v_0$,
the velocity of the string in the $y$-direction
beyond the outward-moving kink.
(One can also easily see that in the laboratory
frame in which the gravitating object is at rest,
the component of the kink velocity
in the $x$-direction is $\cos{\delta}$,
but when one multiplies this by
$\gamma_0 = 1/\sqrt{1-v_0^2} = \sec{\delta}$,
one gets that the kink also moves
at the speed of light along the
distant part that is parallel to the $x$-axis
in the frame of that distant part of the string.)

	Therefore, for a string marginally
captured by a Reissner-Nordstrom black hole,
the relation between
the velocity $v_0$ of the distant straight
part of the string and the stationary point
of nearest approach to the black hole,
at Schwarzschildean radial coordinate $r_0 = R_0 + M$,
or at the conformally-flat radial coordinate $\rho_0$, is
 \begin{eqnarray}
 v_0 &=& \sin{\delta} = \cos{[k'K(k)]}
	\nonumber \\
 &=& \cos{\left[ {\sqrt{R_0^2 - M^2 + Q^2}\over R_0} 
	K\left( {\sqrt{M^2-Q^2}\over R_0} \right) \right]}
	\nonumber \\
 &\approx& {\pi \over 8} k^2 + {7\pi\over 128} k^4
	\nonumber \\
 &=& {\pi (M^2-Q^2) \over 8 R_0^2} + {7\pi(M^2-Q^2)^2\over 128 R_0^4}
	\nonumber \\
 &\approx& {\pi (M^2-Q^2) \over 8 \rho_0^2}
	 - {\pi(M^2-Q^2)^2\over 128 \rho_0^4}.
 \label{eq:59}
 \end{eqnarray}
This gives, for small $v_0$
or small $(M^2-Q^2)/\rho_0^2$,
 \begin{equation}
 \rho_0 = \left[ 1 - {v_0\over 4\pi} + O(v_0^2) \right]
	\sqrt{{\pi(M^2-Q^2)\over 8 v_0}},
 \label{eq:60}
 \end{equation} 
which indeed is $\rho_0 = L$ to lowest order in $v_0$.
For a Kerr-Newman black hole,
the lowest-order term in $v_0$
(which goes as its inverse square root)
is independent of the hole angular momentum
and so is as given in Eq. (\ref{eq:60}),
but the next-order term
(which goes as the square root of $v_0$)
will depend upon the square of the
angular momentum,
and upon the hole orientation
relative to that of the string,
as well as on $M^2-Q^2$,
and so it is not accurately given by Eq. (\ref{eq:60}).

	(The fact that the correction term goes linearly
in $v_0$, rather than quadratically as $\gamma_0$ does,
shows that for strings that become marginally trapped
by the gravitating object at large $\rho_0 \gg M$,
the initial velocity $v_0$ must be so low,
and $\gamma_0$ is so close to one,
that for such strings it is meaningless to include
the $\gamma_0$ factor in the definition (\ref{eq:49})
for the length scale $L$.
However, it is included there to show
how the effect of the $M^2-Q^2$ term in the metric
depends on $v_0$ at all $v_0$,
including values of $v_0$ very near unity or
$\gamma_0 \gg 1$, but still $\gamma_0 \ll b/M$
so that the string stays approximately straight,
as it scatters off the gravitating object without being captured.)

\section{Solution of the motion for very low velocity}

	Now in principle, one can solve
Eqs. (\ref{eq:51})-(\ref{eq:52})
algebraically for $y'(y,z)$ and $z'(y,z)$,
(\ref{eq:53})-(\ref{eq:54})
for $r'(r,\theta)$ and $\theta'(r,\theta)$,
(\ref{eq:55})-(\ref{eq:56})
for $r'(r,c)$ and $c'(r,c)$, or
(\ref{eq:57})-(\ref{eq:58})
for $z'(z,c)$ and $c'(z,c)$,
and then get the autonomous equation
 \begin{equation}
 {dz \over dy} \approx {z'(y,z) \over y'(y,z)},
 \label{eq:61}
 \end{equation}
 \begin{equation}
 {dr \over d\theta} \approx {r'(r,\theta) \over \theta'(r,\theta)},
 \label{eq:62}
 \end{equation} 
 \begin{equation}
 {dr \over dc} \approx {r'(r,c) \over c'(r,c)},
 \label{eq:63}
 \end{equation}
or 
 \begin{equation}
 {dz \over dc} \approx {z'(z,c) \over c'(z,c)}.
 \label{eq:64}
 \end{equation} 
The integral curve that matches the boundary condition
$z=b$ at $y=-\infty$ or $c=-1$ then gives the trajectory
in the $y-z$ plane of the point
on the string ($x=0$) closest to the gravitating object.

	However, solving algebraically for
the derivatives with respect to $y_0 = v_0 t_0$
always involves cubic equations whose
explicit solutions are rather messy,
and the resulting autonomous equations
will almost certainly not be exactly separable
in terms of elementary functions,
so a different technique will employed,
expanding in the dimensionless parameter
$\alpha$ defined in Eq. (\ref{eq:50}).

	In particular, we start with the zeroth-order
solution in $\alpha$ by noting that
when we drop the terms proportional to $\alpha$
in Eqs. (\ref{eq:57})-(\ref{eq:58}),
and take their ratio, we get
 \begin{equation}
 {dz \over dc} \simeq - {L^2 \over z}.
 \label{eq:65}
 \end{equation} 
(Here and henceforth, I shall use $\simeq$
for the approximation that ignores
correction terms of order $\alpha$,
whereas I use $\approx$
for the better approximation that one is
far from the gravitating object in units of its size,
and that the string is at each moment of time nearly straight.)

	With the boundary condition that $z=b$ at $c=-1$,
this simple differential equation has the solution
 \begin{equation}
 z^2 \simeq b^2 - 2 L^2 (1+c) \equiv b^2 - 2 L^2 (1+\cos{\theta})
	\equiv b^2 - 4 L^2 \sigma^2,
 \label{eq:66}
 \end{equation} 
where
 \begin{equation}
 \sigma \equiv \sin{\left( {\pi - \theta \over 2} \right)}
	= \cos{\left( {\theta \over 2} \right)}  = \sqrt{{1+c\over 2}}
	\equiv \sqrt{{1+\cos{\theta}\over 2}}.
 \label{eq:67}
 \end{equation} 

	We see from this that if $b > b_{\rm crit} \simeq 2L$,
the trajectory goes from $c=-1$ ($\theta = \pi$ or $\sigma = 0$)
all the way to $c=+1$ ($\theta = 0$ or $\sigma = 1$)
with $z^2$ remaining positive,
so such a string is not captured but rather scatters from
the gravitating object, with $z^2$ decreasing by an amount
that is roughly $4 L^2$ (correct to zeroth order in $\alpha$),
from $b^2$ to roughly $b^2 - b_{\rm crit}^2 \simeq b^2 - 4 L^2$.
However, for $b < b_{\rm crit} \simeq 2L$,
$z^2$ approaches 0 at
$c \equiv \cos{\theta} \simeq 2 b^2/b_{\rm crit}^2 - 1$,
$\sigma \simeq b/b_{\rm crit}$, or
 \begin{equation}
 \theta \simeq 2 \arccos{b\over b_{\rm crit}}.
 \label{eq:68}
 \end{equation} 
This means that the string is captured by the gravitating object
when it has an impact parameter $b < b_{\rm crit} \simeq 2L$.

	During the final capture process by a black hole,
when the string gets down to $r$ of the order of $M$,
the string will no longer be nearly straight,
so the trajectory of the nearest point on the string
to the black hole will no longer be given
by all of the equations above.
However, then this point will just fall
in nearly radially, so its path in the $y-z$ plane,
though not the $y_0$-dependence of this path,
will still be given approximately correctly by
Eq. (\ref{eq:66}).
For capture by other gravitating objects,
the trajectory will differ when the string
gets sufficiently near or within
the object that the spacetime metric
is no longer well approximated
by the weak-field portion of the
Reissner-Nordstrom metric Eq. (\ref{eq:1}).

	In the intermediate case in which $b = b_{\rm crit} \simeq 2L$,
the point on the string nearest the gravitating object will,
asymptotically with large $t_0$ and with large $t$, approach
the unstable fixed point which is approximately at $(y,z) = (L,0)$,
or, more precisely, at $z=0$ and at $y=\rho_0$
with $\rho_0$ given by the solution of Eq. (\ref{eq:59}).
In this special case, one can integrate $dy_0/dc = 1/c'$
explicitly in terms of elementary functions of $c$
(as opposed to elliptic integrals needed
for a generic impact parameter).
The results are slightly simpler in terms of $\sigma$
defined in Eq. (\ref{eq:67})
(the sine of half the angle from the negative $y$-axis,
which thus goes from 0 when $y=-\infty$
to 1 when the string asymptotically approaches
the fixed point on the positive $y$-axis),
and there is an even simpler expression in terms
of a new coordinate
 \begin{equation}
 \psi \equiv {\rm gd}^{-1} \left( {\pi - \theta \over 2} \right)
	= {1\over 2} \ln{1 + \cos{(\theta/2)} \over 1 - \cos{(\theta/2)}},
 \label{eq:69}
 \end{equation} 
the inverse gudermannian of half the angle $\pi - \theta$
from the negative $y$-axis
(not to be confused
with the previous use of $\psi$ to denote
an exponent in a metric component),
which goes from 0 to $\infty$ and is defined so that
 \begin{equation}
 \sigma \equiv \sin{\left( {\pi - \theta \over 2} \right)} = \tanh{\psi}.
 \label{eq:69b}
 \end{equation} 
Then for the string which approaches with
the critical impact parameter $b \simeq 2L$, one gets
 \begin{eqnarray}
 y_0 &\simeq & {L\over 2} \ln{\left( {1+\sigma \over 1-\sigma} \right)}
	- {L \over \sigma} = L (\psi - \coth{\psi}),
 \label{eq:70} \\
 t &\simeq & t_0 = {y_0\over v_0}
	\simeq {L\over 2 v_0} \ln{\left( {1+\sigma \over 1-\sigma} \right)}
	- {L \over v_0 \sigma} = {L \over v_0} (\psi - \coth{\psi}),
 \label{eq:71} \\
 r &\simeq & {L \over \sigma} = L \coth{\psi},
 \label{eq:72} \\
 y &\simeq & L \left( {-1+ 2\sigma^2 \over \sigma} \right)
	= L (2\tanh{\psi} - \coth{\psi}) \simeq {2L^2 \over r} - r,
 \label{eq:73} \\
 z &\simeq & 2L \sqrt{1-\sigma^2} \ = \ 2L \: {\rm sech}\,\psi
	\ \simeq \ {2L \, \sqrt{r^2 - L^2} \over r}.
 \label{eq:74}
 \end{eqnarray}

\section{Discontinuities in the critical impact parameter}

	For $v_0 \not\ll 1$, where my analytic solution above
of the critical capture is not valid, one would still have
that the critical capture has the point on the string
asymptotically approach the unstable fixed point
at $z=0$ and at $y=\rho_0$
with $\rho_0$ given by the solution of Eq. (\ref{eq:59}).
However, the relation between $b_{\rm crit}$
and $\rho_0$ would not be so simple as the one above,
$b_{\rm crit} \simeq 2 \rho_0$ for $v_0 \ll 1$.
For example, for a Schwarzschild black hole ($Q=0$)
and for $v_0 = 1$, Eq. (\ref{eq:59}) gives $R_0 = M$,
and then Eq. (\ref{eq:9c}) shows that $\rho_0 = M/2$.
On the other hand, $b_{\rm crit} = 3\sqrt{3} \; M$
at this limit of the speed of light \cite{JPD},
so there $b_{\rm crit} = 6\sqrt{3} \; \rho_0$,
with the ratio being $3\sqrt{3}$ times larger
at the speed of light than at very low velocities.

	To get a ratio that does not vary quite so much,
one might instead use the Schwarzschildean
or circumferential radial coordinate
of the unstable fixed point,
which by Eq. (\ref{eq:9c}) is
 \begin{equation}
 r_0 = \rho_0 \left( 1 + {M\over 2 \rho_0} \right)^2 - {Q^2\over 4\rho_0}.
 \label{eq:9r}
 \end{equation}
For large $\rho_0/M$ ($v_0 \ll 1$), $r_0 \sim \rho_0$,
but for $\rho_0 = M/2$ ($v_0 = 1$), $r_0 = 2M = 4 \rho_0$
for a Schwarzschild black hole.
Then one could define the ratio
 \begin{equation}
 F(v_0) \equiv {b_{\rm crit} \over 2 r_0},
 \label{eq:9s}
 \end{equation}
which goes from 1 at $v_0 = 0$
to $(3/4)\sqrt{3} \approx 1.299038$ at $v_0 = 1$
for a test string in the Schwarzschild geometry.

	An interesting point about $F(v_0)$,
which is a dimensionless number of order unity
(a function of the string initial velocity $v_0$,
and of $(Q/M)^2$ for a Reissner-Nordstrom black hole)
that parametrizes the critical impact parameter
for the capture of a infinite test string by a black hole,
is that I would expect it
(and hence also the critical impact parameter $b_{\rm crit}$,
since $r_0$ is a continuous function of $v_0$)
to have an infinite number
of discontinuities as $v_0$ is increased from 0 to 1.
These discontinuities should arise
from the discreteness of the number
of times the closest point on the string (at $x=0$)
wraps around the black hole before
this point approaches the unstable fixed point
characterizing the critical capture solution
at a particular $v_0$.
At very low $v_0$ we found above that
the angle $\theta$ of this closest point on the string
from the positive $y$-axis
decreased monotonically from $\pi$ down to 0
in approaching the unstable fixed point,
but for a sufficiently larger values of $v_0$,
I would expect it to decrease to $-2\pi n$
for some positive integer $n$ that depends
discontinuously on $v_0$
and characterizes how many
times the string wraps around the black hole
before it makes its final asymptotic approach
to the unstable fixed point.
As $v_0$ approaches 1,
$n$ should approach $\infty$,
as the string wraps around the black hole
an infinite number of times before
approaching the unstable fixed point
characterizing the final marginally bound string.

	Unless there are some surprising cancellations,
I would expect that $F(v_0)$, and hence
$b_{\rm crit}(v_0) = F(v_0) r_0(v_0)$,
to have a discontinuity each time
$v_0$ crossed a discontinuity
of the wrapping integer $n$,
with an infinite number of such discontinuities
occurring as one approaches the speed of light
($v_0 = 1$).
Indeed, when I voiced this prediction,
De Villiers \cite{JPDprivate}
confirmed that he has seen a suggestion
of it in his numerical calculations \cite{JPD3}.

	The situation would also be complicated
by the fact that cusps and loops form
in the string when $v_0$ is large enough \cite{Moss}.
To get a simple unique definition for $b_{\rm crit}(v_0)$,
one could assume that any loops that form
do not break off, in which case it seems clear
that the marginal capture behavior
is indeed when the point on the string
approaches the unstable fixed point.
However, if one allowed loops to break off
the string, one would have to specify
when the loops break off and whether
capture was defined to be the case in
which at least one such loop ended up
orbiting or falling into the black hole,
or whether the piece connected to infinity
itself has to become bound to the hole.
In the former specification I am not
certain that one would necessarily
get a critical impact parameter that
is discontinuous in the velocity
(as I am arguing that it is when one
does not allow loops to break off,
and which it also should be if capture
is defined so that the piece of the string
connected to infinity
itself has to become bound to the hole).
However, it is not obvious to me how
to come up with a definition
that would make the critical impact parameter
a continuous function of the string
initial velocity $v_0$.
In any case in which loops of string break off,
they would carry away momentum and
alter the critical impact parameter
in a way that would depend upon
when the loops break off. 

\section{First-order corrections at finite velocity}

	Now return to the analysis of low initial string
velocities, $v_0 \ll 1$, where the wrapping integer $n$ is zero
and the critical impact parameter $b_{\rm crit}(v_0)$
is a continuous function of $v_0$.
Let us evaluate the first-order corrections
in the dimensionless parameter $\alpha$ defined
by Eq. (\ref{eq:50}).  Define the quantity
 \begin{equation}
 B \equiv {z^2 \over 2L^2} + c = {z^2 \over 2L^2} + 2\sigma^2 - 1
	\simeq {b^2 \over 2 L^2} - 1,
 \label{eq:75}
 \end{equation} 
which to zeroth order in $\alpha$ is thus constant.
From Eqs. (\ref{eq:51})-(\ref{eq:58}),
one can deduce that the derivative of $B$
with respect to $c \equiv \cos{\theta} \equiv 2\sigma^2 - 1$ is
 \begin{equation}
 {dB \over dc} \approx - \alpha \left( {r\over L} y' - {L\over r} r' \right).
 \label{eq:76}
 \end{equation} 
Evaluate $r'$ and $y'$ to zeroth order in $\alpha$ to get
 \begin{equation}
 {dB \over dc} \cong
 - \alpha \left( {r\over L} - {2Lc\over r} + {L^3\over r^3} \right),
 \label{eq:77}
 \end{equation} 
where the $\cong$ sign means that the equation
is accurate through first order in $\alpha$
(and thus better than the $\simeq$ approximations,
which are only accurate to zeroth order in $\alpha$,
but not so good as the $\approx$ approximations,
which are accurate for all $\alpha$ and which essentially only
require that the string configurations be nearly straight).

	Next, insert the zeroth-order solution for $r(c)$
into the right hand side of Eq. (\ref{eq:77}).
To simplify the result,
replace the independent variable $c$ with $\sigma$
and define the dimensionless parameter
 \begin{equation}
 k \equiv {b_{\rm crit}\over b} \simeq {2L\over b},
 \label{eq:78}
 \end{equation}
not to be confused with the previous
use of $k$ as defined in Eq. (\ref{eq:6});
I use the same letter because this $k$,
like the previous one, will be used
as the modulus of a complete elliptic integral.
Then one gets
 \begin{equation}
 {dB \over d\sigma} \cong
 - {4\alpha (1 - 6k^2\sigma^4 + 4k^2 \sigma^6 + k^4\sigma^8)
	\over k (1 - k^2\sigma^2)^{3/2} \sqrt{1-\sigma^2}}.
 \label{eq:79}
 \end{equation} 

	Consider the case in which $k < 1$,
so that the impact parameter $b$ is greater than
the critical impact parameter $b_{\rm crit} \simeq 2L$
and so that the string scatters without being
captured, with $z$ remaining positive
as $\sigma = \cos{\theta/2}$ goes from 0 at $\theta = \pi$
and $y = -\infty$ to 1 at $\theta = 0$ and $y = +\infty$.
Then integrating Eq. (\ref{eq:79}) from $\sigma = 0$
to $\sigma = 1$ gives the change in $B$ in terms of
the complete elliptic integrals $K(k)$ (of the first kind)
and $E(k)$ (of the second kind) with modulus $k$:
 \begin{equation}
 \Delta B \cong {32\alpha\over 15 k^5}
	[(2 - 3k^2 + k^4)K(k) + (-2 + 2k^2 - 2k^4)E(k)].
 \label{eq:80}
 \end{equation}
One can then use this result to get the final
value of $z^2$ as $y$ goes to $+\infty$:
 \begin{equation}
 z_f^2 = b^2 - 4L^2 + 2L^2 \Delta B.
 \label{eq:81}
 \end{equation}

	When the impact parameter is very near
the critical impact parameter, the complementary modulus
 \begin{equation}
 k' \equiv \sqrt{1-k^2} \equiv \sqrt{1-b_{\rm crit}^2/b^2}
 \label{eq:82}
 \end{equation} 
is very small, and from the form of $K(k)$ and $E(k)$
for small $k'$, one gets from Eq. (\ref{eq:80})
 \begin{equation}
 \Delta B \cong -{64\alpha\over 15}
 \left[1 + {5\over 4}k'^2 - O\left(k'^4 \ln{{\rm const.}\over k'} \right) \right].
 \label{eq:83}
 \end{equation}
Taking the limit of this when the impact parameter
reaches the critical impact parameter, so that $k' = 0$,
and inserting the result back into Eq. (\ref{eq:81})
with $z_f = 0$ gives the critical impact parameter
to first order in $\alpha$:
 \begin{equation}
 b_{\rm crit} \cong 2L + {32\over 15} \alpha L
 \approx \sqrt{\pi(M^2-Q^2)\over 2v_0} + {64\over 15} M v_0.
 \label{eq:84}
 \end{equation}
Here I have dropped the factors of $\gamma_0$,
since these expressions are only valid for
$b_{\rm crit} \gg M$, and this implies that $v_0 \ll 1$
so that $\gamma_0 \approx 1$.

	It is amusing that if one evaluates the last
expression on the right hand side of Eq. (\ref{eq:84})
for a Schwarzschild black hole ($Q=0$) at $v_0 = 1$
(which is far outside its range of validity,
which is for $v_0 \ll 1$), one gets
$(\sqrt{\pi/2} + 64/15) M \approx 5.519981 \: M$,
which is not too far from the actual \cite{JPD}
$b_{\rm crit} = 3\sqrt{3}\: M \approx 5.196152 \: M$
at the speed of light.
Since Eq. (\ref{eq:84}) is only valid for
$b_{\rm crit} \gg M$,
it probably misses a $v_0$-independent term
of order $M$.
It is therefore tempting to add such a term
to give the right answer at $v_0 = 1$
and conjecture that a crude approximation
for the critical impact parameter for a straight
string moving at any initial velocity $v_0$
in the field of a Schwarzschild black hole is
 \begin{eqnarray}
 b_{\rm crit} 
 &{\ \lower-1.2pt\vbox{\hbox{\rlap{?}
 \lower5pt\vbox{\hbox{$\!\!\!\sim$}}}}\ }&
 \left[\sqrt{\pi\over 2v_0}
 - \left( \sqrt{\pi\over 2} + {64\over 15} - \sqrt{27} \right)
 + {64\over 15} v_0 \right] M
	\nonumber \\
 &\approx& (1.253314 v_0^{-1/2} - 0.323828 + 4.266667 v_0) \: M.
 \label{eq:85}
 \end{eqnarray}
Here the
${\ \lower-1.2pt\vbox{\hbox{\rlap{?}
 \lower5pt\vbox{\hbox{$\!\!\!\sim$}}}}\ }$
emphasizes that this
is only a very crude guess that gives a good approximation
at both $v_0 \ll 1$ and at $v_0 = 1$,
but it has by no means been derived
at intermediate velocities,
and in this paragraph the $\approx$
just means the numerical approximation
of using a finite decimal approximation
for the numbers given exactly before the $\approx$.
Of course, this crude guess is also continuous and so
misses the infinite number of discontinuities
that were predicted above for $b_{\rm crit}(v_0)$.

	It is also amusing to note that this crude guess
(\ref{eq:85}) gives a minimum critical impact parameter of
 \begin{eqnarray}
 b_{\rm crit \; min}
 &{\ \lower-1.2pt\vbox{\hbox{\rlap{?}
 \lower5pt\vbox{\hbox{$\!\!\!\sim$}}}}\ }&
 \left[ 6 \left({\pi\over 15}\right)^{1/3}
 - \sqrt{\pi\over 2} - {64\over 15} + \sqrt{27} \right] M
	\nonumber \\
 &\approx& 3.239349 \: M
 \label{eq:85b}
 \end{eqnarray}
at $v_0 {\ \lower-1.2pt\vbox{\hbox{\rlap{?}
 \lower5pt\vbox{\hbox{$\!\!\!\sim$}}}}\ }
 (225\pi)^{1/3}/32 \approx 0.278373$.
Numerical calculations \cite{JPD}
do show that the critical impact parameter
decreases as $v_0$ is reduced below unity,
and they hint of the rise at lower impact parameters
that must occur for Eq. (\ref{eq:84}) to be valid there,
but it remains to be seen how accurate
the crude guess of Eq. (\ref{eq:85b})
is for the minimum critical impact parameter
and the velocity $v_0$ at which it occurs.

	I should warn that although Eq. (\ref{eq:84})
attempts to give the term in the critical impact
parameter that is linear in $v_0$,
I would suspect that this term,
even with a $v_0$-independent term
that I suspect may be present
but which I do not know how to evaluate,
does not give the leading correction
to the dominant term that goes as $v_0^{-1/2}$.
We have seen from Eq. (\ref{eq:60})
that the location of the unstable
critical point for a string that is just
marginally captured by a black hole
has a correction term to $L$
that goes as $v_0^{1/2}$,
and I would suspect that $b_{\rm crit}$
also has a correction term going as $v_0^{1/2}$,
though I do not know what it is.
Therefore, I suspect that for low velocities $v_0$,
the critical impact parameter
for a Schwarzschild black hole may be expanded as
 \begin{equation}
 b_{\rm crit}
 = [\sqrt{\pi/2} \; v_0^{-1/2}
 + O(1) + O(v_0^{1/2}) + O(v_0) + O(v_0^{3/2})] M.
 \label{eq:86}
 \end{equation}
The only thing I can assert with confidence is that
I have derived the coefficient of the leading term,
but I do not know the coefficients of
the $O(1)$ and $O(v_0^{1/2})$ terms
that I suspect are present.
Eq. (\ref{eq:84}) might give the correct $O(v_0)$ term,
but I am also not certain that there are no other
$O(v_0)$ terms that have been missed in the approximations
that I have used.
In other words, the only firm conclusion that
I can make from the approximations that I have made
is that in the limit of low velocity $v_0$
for a straight test string impinging upon a Kerr-Newman black hole,
or for any other gravitating object that is very small
in comparison with the string impact parameter,
 \begin{equation}
 \lim{_{v_0 \rightarrow 0} v_0^{1/2} b_{\rm crit}}
 = \sqrt{\pi(M^2-Q^2)\over 2} .
 \label{eq:87}
 \end{equation}

	Next, go back to Eq. (\ref{eq:80})
and evaluate it when the impact parameter $b$
is much greater than the critical impact parameter
$b_{\rm crit}$, so that $k \equiv b_{\rm crit}/b \ll 1$.  Then
 \begin{equation}
 \Delta B \cong - {2\pi\alpha\over k} + {\pi\alpha k\over 2}
 \cong - \pi\alpha\left( {b\over L} - {L\over b} \right).
 \label{eq:88}
 \end{equation}
However, I suspect that my approximations
have missed out terms that are larger
than the second term of the right hand side of Eq. (\ref{eq:88}),
which is much smaller than the first term,
so I shall henceforth drop the second term.

	Inserting the resulting $\Delta B$
from Eq. (\ref{eq:88}) for $b \gg 2L$
into Eq. (\ref{eq:81}), and taking the
square root to the same level of approximation, gives
the final height of the string, over the plane
through the gravitating object location that is parallel
to the string's original extension and motion, as
 \begin{equation}
 z_f \cong b - {2L^2\over b} - \pi\alpha L
 \approx b - {\pi(M^2-Q^2)\over 4v_0 b} - 2\pi M \gamma_0 v_0.
 \label{eq:89}
 \end{equation}
Here I have dropped the $\gamma_0$ factor
from the second term,
since it is only significant in comparison
with the third term if $v_0 \ll 1$, which gives
$\gamma_0 \approx 1 + v_0^2/2 \approx 1$.
However, I have retained the $\gamma_0$
factor in the third term,
since it is actually valid for all values
of the celerity $\gamma_0 v_0$.
This formula is precisely the same as that derived
independently in \cite{JPDnew}.

	One might think that Eq. (\ref{eq:89}) is only valid
for $\alpha \ll 1$, which one can easily see from
the definition of $\alpha$ in Eq. (\ref{eq:50})
implies that $v_0 \ll 1$.
Indeed, my derivation did assume this
when I expanded in powers of $\alpha$
and kept only the lowest nontrivial power.
However, the leading term in $\Delta B$
at large $b/L$, actually comes from just
the first term of Eq. (\ref{eq:76}),
which is then simply
 \begin{equation}
 {dB \over dc} \approx - \alpha \left( {r\over L} y' \right),
 \label{eq:90}
 \end{equation}
without using any approximation that $\alpha \ll 1$. 
Continuing at large $b/L$, for the right hand side
it is a sufficiently good approximation to
set $z \approx b$
(so $r \approx b/\sin{\theta} = b/\sqrt{1-c^2}$
and $y' \approx 1$), in which case Eq. (\ref{eq:90})
becomes
 \begin{equation}
 {dB \over dc} \approx - {\alpha b\over L \sqrt{1-c^2} },
 \label{eq:91}
 \end{equation}
which, when integrated from $c=-1$ to $c=1$, gives
 \begin{equation}
 \Delta B \approx - {\pi\alpha b\over L},
 \label{eq:92}
 \end{equation}
independent of the size of $\alpha$.
This leads to Eq. (\ref{eq:89})
for all values of $v_0$, so long as
$b \gg b_{\rm crit} \simeq 2L$.

	Even more simply,
one may derive Eq. (\ref{eq:89})
directly from Eq. (\ref{eq:52})
by putting $y \approx y_0$
and $z \approx b$
(and hence $y' \approx 1$
and $z' \approx 0$)
on the right hand side
and integrating.

	By using these results for
$b$ near $b_{\rm crit}$
and for $b\gg b_{\rm crit}$,
we can replace Eq. (\ref{eq:81}),
with its $\Delta B$ given by Eq. (\ref{eq:80})
in terms of complete elliptic integrals,
with a simpler explicit expression
that has almost as much relative accuracy
for $z_f^2$ (except when it is very tiny)
and for $z_f^2 - b^2$,
though some relative accuracy is lost
for the difference between $z_f^2$
and its zeroth-order (in $\alpha$)
approximation $b^2 - 4 L^2$
when this difference is small, which occurs
when $b$ is not much larger than $2 L$.
Namely, we may write
 \begin{equation}
 z_f^2 \approx
 b^2 - {\pi(M^2-Q^2)\over 2 v_0} - 4\pi M \gamma_0 v_0 b,
 \label{eq:93}
 \end{equation}
which is an excellent approximation
so long as $b \gg M \gamma_0$ and so long as
the resulting $z_f^2$ it gives
is positive and much larger than
the magnitude of the third (last)
term on the right hand side.

	To see that this is very nearly what
the more accurate Eq. (\ref{eq:81}) gives,
we can consider various cases.
First, consider the case in which
the magnitude of the second term
on the right hand side of Eq. (\ref{eq:93})
is larger than the magnitude of the third term.
This occurs for $b < (M^2-Q^2)/(8M\gamma_0 v_0^2)$.
Since we are requiring $M \ll b$,
this case implies that $v_0 \ll \sqrt{1-Q^2/M^2}$,
a low velocity, justifying the omission
of the $\gamma_0$ factor that is in the
second term of the right hand side
of Eq. (\ref{eq:81}) ($-4 L^2$) but which is dropped
in the simplified second term of Eq. (\ref{eq:93}).

	Then the third term of the right hand side
of Eq. (\ref{eq:81}) ($+ 2L^2 \Delta B$),
which is always of the same order of magnitude
as the third term of the right hand side
of Eq. (\ref{eq:93}) ($ - 4\pi M \gamma_0 v_0 b$)
but which differs by a factor of order unity
that is significantly different from unity
when $b$ is not very large
compared with $b_{\rm crit} \sim 2 L$,
will be much less than
the second term unless $b$ nearly saturates
the inequality given above.
If $b$ does nearly saturate this inequality,
one can easily see that one must have
$b \gg b_{\rm crit}$.
Therefore, when the modulus $k$
of the elliptic functions in Eq. (\ref{eq:80})
for $\Delta B$ is of order unity
($b \not\gg b_{\rm crit}$),
the third term of the right hand side of
Eq. (\ref{eq:81}) ($+ 2L^2 \Delta B$), and
the third term of the right hand side of
Eq. (\ref{eq:93}) ($ - 4\pi M \gamma_0 v_0 b$),
are both very small compared with the second term,
and so they may both be ignored,
even though they may differ by
a factor of order unity
(e.g., by a factor of about $15\pi/16$ at $b = b_{\rm crit}$).
On the other hand,
when the third terms of the right hand sides of
Eqs. (\ref{eq:81}) and (\ref{eq:93})
are not negligible in comparison
with the second term,
one has $b \gg b_{\rm crit}$
and hence $k \ll 1$,
so that these two third terms
are very nearly the same.

	In the opposite case, in which
the magnitude of the second term
on the right hand side of Eq. (\ref{eq:93})
is smaller than the magnitude of the third term,
which occurs for $b > (M^2-Q^2)/(8M\gamma_0 v_0^2)$,
by invoking our assumption $b \gg M\gamma_0$
one can easily see that in this case
$k^2 \approx 4L^2/b^2 = \pi(M^2-Q^2)\gamma_0/(2 v_0 b^2) \ll 1$,
so $\Delta B$ is well approximated by $-\pi\alpha b/L$
and the third terms of the right hand sides of
Eqs. (\ref{eq:81}) and (\ref{eq:93})
are very nearly the same.
 
	In other words, the third term
of the right hand side of Eq. (\ref{eq:93})
is not always an accurate approximation
to the third term of the right hand side
of Eq. (\ref{eq:81}),
but it is when either of these two terms
makes any significant difference to
$z_f^2 - b^2$.

	Eq. (\ref{eq:93}) thus gives a good approximation
for the final height $z_f$ (above the $z=0$ plane
containing the center of mass of the gravitating object
and parallel to the string's initial straight extent
in the $x$-direction and motion with speed $v_0$
in the $y$-direction) in terms of the initial height $z_i = b$,
the speed $v_0$, and the mass $M$ and charge $Q$
of the gravitating object,
so long as the string stays in the electrovac weak-field
region outside the object, far enough away
that one can ignore the metric deviations
from sphericity due to the angular momentum
and multipole moments of the gravitating object,
and so long as $b \gg M\gamma_0$
so that the third term of Eq. (\ref{eq:93})
is small compared with its first term ($b^2$).

\section{Formulas for arbitrary velocity}

	It would be nice to remove the
restriction to $\gamma_0 \ll b/M$
(which is implied by the method used above
in order that the string remain nearly straight
during its entire scattering),
and indeed one can do so
by trivially extending \cite{Page_fast}
to arbitrarily large celerity
the results of \cite{JPDnew},
that to linear order in $M$
(i.e., ignoring the $M^2-Q^2$ terms
that are only negligible
when $v_0^2 \gg (M/b)(1-Q^2/M^2)$),
the final deflection is simply
$z_f - b \sim - 2\pi M \gamma_0 v_0$
(assuming that the string
stays in the far-field region where
the gravitational field is not only
weak, $b \gg M$,
but also nearly spherically symmetric,
and assuming that any loops which
may form in the string do not break off
to carry away momentum).

	The argument that this formula
is valid for arbitrarily large celerities
(and hence arbitrarily large $\gamma_0$,
even for $\gamma_0 > b/M$)
is, in brief, the following
\cite{Pagenon,Page_fast}:
First, if one has an infinitely long string
of tension $\mu$
passing far from a gravitating object
($b \gg M/v_0^2$,
and also $b v_0^2$ much larger
than the scale given by any distortions
from the spherical electrovac
Reissner-Nordstrom metric),
in the test string approximation
$\mu \ll M \gamma_0 v_0/b$ \cite{Pagenon}
so that the gravitating object does
not accelerate significantly
from the back reaction of the string,
then to first order in $M$
(which dominates when the inequalities
above hold),
the momentum transferred
from a gravitating object to the string is
$4 \pi \mu M \gamma_0 v_0$,
perpendicular to the motion of the string
and toward the side on which
the object passes the string,
for arbitrarily large $\gamma_0$.
Second, it is easy to see \cite{Pagenon,Page_fast}
that an infinitely long, initially straight,
test string in flat spacetime
undergoes a total transverse displacement
from its otherwise uniform motion
that is simply the transverse momentum transferred
to the string divided by $2\mu$,
whether or not cusps or loops form
(so long as they do not break off
to carry away momentum
from the infinitely long connected
portion of the string).

	One can easily see that this formula
for the final deflection is indeed
what Eq. (\ref{eq:93}) gives when
one ignores the second term
(with the $M^2-Q^2$ factor)
and takes $b \gg M \gamma_0$
so that the third term is small
in comparison with the first term.
But when one does not have
$b \gg M \gamma_0$,
Eq. (\ref{eq:93}) is not accurate.
E.g., if the third term is larger
than the first term, it would
nonsensically give a negative
$z_f^2$, even though the string
scatters and is not captured as it would
be if the second term were larger
than the first term
(when a negative expression for $z_f^2$
does have the meaning that something
went wrong with the assumption
that the string scattered).

	However, one can easily combine
Eq. (\ref{eq:93}), which is valid for $b \gg M \gamma_0$,
with $z_f - b \sim - 2\pi M \gamma_0 v_0$ \cite{JPDnew},
which is valid for $b \gg M/v_0^2$
for arbitrarily large $v_0$ \cite{Page_fast},
as briefly argued above, to get
 \begin{equation}
 z_f \ \approx \ 
 b \; \sqrt{ 1 - {\pi(M^2-Q^2)\over 2 b^2 v_0}} \; - \; 2\pi M \gamma_0 v_0,
 \label{eq:94}
 \end{equation}
which is valid for all test string velocities $v_0$
so long as $b \gg M$ so that the string
stays in the weak-field regime
(and also assuming that the gravitational
field is electrovac and nearly spherically symmetric
in this regime).

	Therefore, by devising an approximation
that works for all infinitely long test strings
that pass by a gravitating
object at sufficiently large distance that
that stay nearly straight at each moment of time
that they are influenced significantly 
by the gravitational field of the object,
we can extend the analysis of \cite{JPDnew}
to arbitrarily low and high velocities.  We thereby find
the critical impact parameter for capture
at low velocity,
 \begin{equation}
 b_{\rm crit} = M \left( \sqrt{(\pi/2)(1-Q^2/M^2)/v_0} \: 
 + O(1) + O(v_0^{1/2}) + {64\over 15} v_0 \right)
 \label{eq:95}
 \end{equation}
(though I have not ruled out the possibility
of other $O(v_0)$ terms that could change
the coefficient $64/15$),
and we find that if the actual impact parameter, $z_0 = b$,
is greater than the critical impact parameter $b_{\rm crit}$,
the final height of the string, $z_f$, is given
to a good approximation by Eq. (\ref{eq:94})
for all velocities $v_0$ when $b \gg M$.

\section*{Acknowledgments}

	I thank Jean-Pierre De Villiers and Valeri Frolov
for pointing out to me that the gravitational capture
of slowly moving test strings was an unsolved problem,
and, shortly after my calculations were completed
and written up, for giving me their paper \cite{JPDnew}
which independently derived the perturbative
scattering of strings and which agrees with
the limiting case of my results in which one takes
the impact parameter to be much larger than
the critical impact parameter for capture.
I also thank them for further discussions, for
guiding me to some of the literature on this subject,
and for comments on my manuscript.
This work was supported in part by the
Natural Sciences and Engineering Research Council
of Canada.

\end{document}